\documentclass[lettersize,journal]{IEEEtran}
\usepackage[T1]{fontenc}

\usepackage{subfigure}
\usepackage{colortbl}
\usepackage{bm}
\usepackage{fancyhdr}       

\usepackage{graphicx}  
\usepackage{url}       

\usepackage{amsmath}   
\usepackage{cite}

\usepackage{amsfonts,amssymb}

\usepackage{stfloats}
\usepackage{indentfirst} 
\usepackage{cases}
\usepackage{comment}
\usepackage{algorithm}
\usepackage{multirow}
\usepackage{algorithmic}

\usepackage{multirow}
\usepackage{graphicx}
\usepackage{epstopdf}
\usepackage{diagbox}
\usepackage{subfig}
\usepackage{caption}
\usepackage{tikz}
\usepackage{color}

\usepackage{algorithm}
\usepackage{algorithmic}
\usepackage{amsmath}

\usepackage{graphicx}
\usepackage{epstopdf}

\usepackage{graphicx}  
\usepackage{url}       

\usepackage{amsmath}   
\usepackage{cite}
\usepackage{extarrows}
\usepackage{amsfonts,amssymb}

\usepackage{stfloats}

\usepackage{cases}
\usepackage{algorithm}
\usepackage{multirow}
\usepackage{algorithmic}
\usepackage{graphicx}
\usepackage{epstopdf}

\usepackage{graphicx}  
\usepackage{url}       

\usepackage{amsmath}   
\usepackage{cite}
\usepackage{extarrows}
\usepackage{amsfonts,amssymb}

\usepackage{stfloats}

\usepackage{amsfonts}

\usepackage{cases}
\usepackage{algorithm}
\usepackage{multirow}
\usepackage{algorithmic}

\newcommand{\ls}[1]
    {\dimen0=\fontdimen6\the\font
     \lineskip=#1\dimen0
     \advance\lineskip.5\fontdimen5\the\font
     \advance\lineskip-\dimen0
     \lineskiplimit=.9\lineskip
     \baselineskip=\lineskip
     \advance\baselineskip\dimen0
     \normallineskip\lineskip
     \normallineskiplimit\lineskiplimit
     \normalbaselineskip\baselineskip
     \ignorespaces
    }

\graphicspath{{Visio-File-0731-final/}{Simulation-1213/}}

\begin{document}
\title{{Delay Tradeoff and Adaptive Finite Blocklength Framework for URLLC}}

\author{Yixin Zhang, Wenchi Cheng, Jingqing Wang, and Wei Zhang\vspace{-2pt}
\\

}
\date{\today}


\maketitle

\thispagestyle{empty}

\begin{abstract}
With various time-sensitive tasks to be served, ultra-reliable and low-latency communications (URLLC) has become one of the most important scenarios for the fifth generation (5G) wireless communications. The end-to-end delay from the sub-millisecond-level to the second-level is first put forward for a wide range of delay-sensitive tasks in the future sixth generation (6G) communication networks, which imposes a strict requirement on satisfying various real-time services and applications with different stringent quality of service (QoS) demands. Thus, we need to find out new delay reduction framework to satisfy the more stringent delay requirements. In this article, a state-of-the-art overview of end-to-end delay composition and delay analysis combined with access protocols are elaborated. We reveal the tradeoff relationship among transmission delay, queuing delay, and retransmission times with the change of blocklength in the finite blocklength (FBL) regime. Based on the tradeoff and combining the grant-free (GF) random access (RA) scheme, we propose the adaptive blocklength framework and investigate several effective algorithms for efficiently reducing the over-the-air delay. Numerical results show that our proposed framework and schemes can significantly reduce the over-the-air delay for URLLC. 
\end{abstract}


\setcounter{page}{1}

\IEEEoverridecommandlockouts

\maketitle

\section{Introduction}

\IEEEPARstart{U}{ltra-reliable} and low-latency communications (URLLC), the core business of the fifth-generation (5G) communication networks, aims to provide the end-to-end delay of less than $1$~ms with a $99.999\%$ success probability for $32$-bit packet transmission specified by the 3rd Generation Partnership Project (3GPP)~\cite{38.913}, enabling various low-latency real-time services. The upcoming sixth generation (6G) communication networks are expected to support a wide range of Internet of Everything (IoE) services and applications, such as autonomous driving, remote control, augmented/virtual reality (AR/VR), factory automation, and Tactile Internet, thus putting forward more stringent end-to-end delay requirement with sub-millisecond level for delay-sensitive tasks in the next generation wireless networks~\cite{6G_Walid}.

Most IoE traffics are with short packets, where the finite blocklength (FBL) theory has been proposed as one of the effective methods for short packets to reduce the transmission delay~\cite{ChannelRate}. The FBL theory provides an accurate tool to describe the tradeoff among delay, reliability, and achievable rate for single short packet transmission. Introducing shorter transmission time intervals (TTIs) and using smaller blocklengths in 5G New Radio (NR), the transmission delay of an individual packet can be hence reduced. The joint blocklength and power optimization method for downlink transmission schemes were proposed, aiming to minimize the decoding error probability within the constraints of total energy and delay~\cite{Blocklength_Minimization}. However, end-to-end delay not only includes transmission delay, but also contains queuing delay, processing delay, propagation delay, backhaul delay, routing delay, and the number of retransmissions caused by transmission error probability~\cite{E2E_new}. To further reduce delay to meet the sub-millisecond level delay demand in 6G, it is necessary to comprehensively consider the impact of FBL on each delay component of the end-to-end delay.

In addition, the short TTIs introduced by 5G NR are based on a fixed blocklength structure, which impacts the flexibility of URLLC~\cite{Fixed_TTI}. Firstly, fixed blocklength is difficult to satisfy the diverse quality of service (QoS) needs of IoE, including video streaming, instant messaging, sensor data, etc. Secondly, packet arrival is highly variable and bursty in IoE scenarios, where fixed blocklength cannot be able to effectively adapt to traffic changes, which leads to large delay and low transmission efficiency. Next, some blocks only contain a small amount of data while the rest is idle, resulting in resource waste and low resource utilization. Also, the transmission error probability no longer approaches zero, thereby exacerbating the retransmission problem and increasing delay. Therefore, it is necessary to consider a more flexible blocklength structure, such as variable blocklength, to be more suitable for fast and reliable packet transmission in URLLC. 

Considering the large number of low-latency devices in IoE scenarios, we also need to consider multiple access protocols, which lead to different delay values and retransmission conditions. Thus, we need to analyze the end-to-end delay reduction problem in combination with the access protocol. The grant-based (GB) random access (RA) consists of four-step handshaking, which wastes time and is not feasible for URLLC. To improve the efficiency and simplify access process, the grant request step is 
eliminated in the grant-free (GF) RA, where the handshaking and waiting delay during the resource scheduling phase can be avoided for URLLC~\cite{GB-GF}. Also, since the channels are shared by multiple users, the collision introduced by GF RA also leads to the increase in retransmission times, which inevitably harms the delay performance. 

In general, the majority of current studies focus on the optimization of a single delay component using a fixed blocklength structure and do not consider specific access protocols, which leads to a lack of practical integration with specific RA protocols, joint optimization of multiple end-to-end delay components, and flexible design of blocklength structure. Fortunately, a novel tradeoff relationship exists among transmission delay, queuing delay, and retransmission times in terms of FBL, serving as a guide to the aforementioned problems. To mitigate the combined impact of delay components on end-to-end delay as blocklength varies, the joint delay optimization scheme and flexible blocklength framework are highly demanded. Therefore, we need to comprehensively analyze all related delay components and access protocols to consider the impact of blocklength for the new delay reduction framework.

The rest of the article is structured as follows. We provide a comprehensive review of end-to-end delay with access protocol for low-latency communications. Then, we reveal the new tradeoff in the FBL regime. We also propose the adaptive blocklength framework and the corresponding delay minimization problem. On this basis, we present a case study and show novel delay-related research directions. Finally, the conclusion is given.

\section{The Delay Composition And Access Protocol}
\subsection{The Components Of End-to-end Delay}
\begin{figure*}[htbp]
  \centering
  \includegraphics[scale = 1]{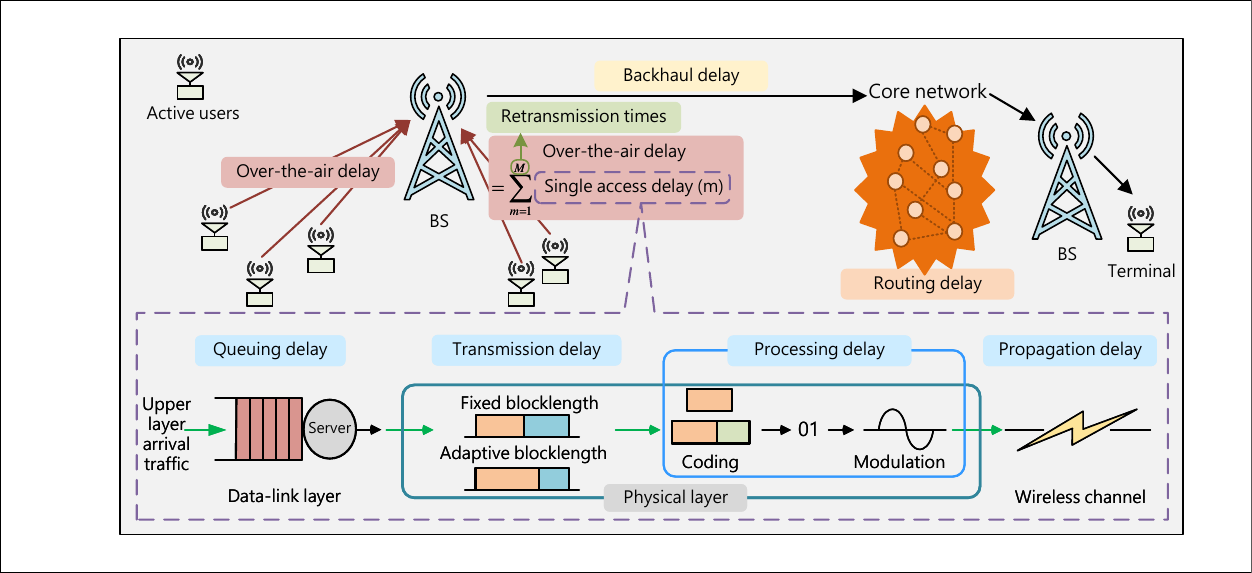}
  \caption{The end-to-end delay components.}\label{System1}
\end{figure*}

Figure~\ref{System1} shows the various end-to-end delay components in URLLC scenario. The user plane end-to-end delay consists of the over-the-air delay, the routing delay in the core networks, and the backhaul delay~\cite{5}. The backhaul delay refers to the time taken for a packet to be transmitted from the core network to the small subnetworks at the edge of the network,  and this duration is influenced by the physical distance from the base station (BS) to the core network as well as the type of physical medium used. The utilization of fiber optics significantly reduces the backhaul delay to a duration of less than $1$ millisecond~\cite{she2017radio}. The routing delay denotes the duration required for a packet to travel through routing paths within a network or across multiple networks. Software-defined routers (SDRs) have become an effective approach to achieve a relatively routing delay while offering programmable capabilities~\cite{RoutingDelay_New}. The over-the-air delay, which is a very important part as well as the bottleneck of end-to-end delay for URLLC, is influenced by the single access delay and the number of access attempts, i.e., retransmission times. Specifically, the single access delay consists of transmission delay, queuing delay, processing delay, and propagation delay. In Table~\ref{Table:Two_Table}, we provide the existing delay reduction technologies of over-the-air delay components. 

\begin{table*}[htbp]
	\centering
	\caption{Reduction technologies of over-the-air delay components and the corresponding access protocol-determined delay.}
	\label{Table:Two_Table}
	\begin{tabular}{|c|cccc|cclclc|}
		\hline
		\multirow{2}{*}{\begin{tabular}[c]{@{}c@{}}Over-the-air delay\\ components\end{tabular}} & \multicolumn{4}{c|}{\multirow{2}{*}{Reduction technologies}}                                                                                                                                                                                                                                                                                                                                                                                                                 & \multicolumn{6}{c|}{Access protocol-determined delay}                                                                                                                                                                                                                                                                                                                                                                 \\ \cline{6-11} 
		& \multicolumn{4}{c|}{}                                                                                                                                                                                                                                                                                                                                                                                                                                                        & \multicolumn{1}{c|}{Description}                                                                & \multicolumn{4}{c|}{GB RA}                                                                                                                                                                              & GF RA                                                                                                     \\ \hline
		\multirow{3}{*}{\begin{tabular}[c]{@{}c@{}}Transmission\\ delay\end{tabular}}            & \multicolumn{1}{c|}{\multirow{3}{*}{Shortened TTIs}}                                                                                    & \multicolumn{3}{c|}{\begin{tabular}[c]{@{}c@{}}Increase the subcarrier spacing (SCS):\\ to reduce the symbol duration.\end{tabular}}                                                                                                                                                                                               & \multicolumn{1}{c|}{\begin{tabular}[c]{@{}c@{}}Number\\ of times\end{tabular}}                  & \multicolumn{4}{c|}{2}                                                                                                                                                                                  & 1                                                                                                         \\ \cline{3-11} 
		& \multicolumn{1}{c|}{}                                                                                                                   & \multicolumn{3}{c|}{\begin{tabular}[c]{@{}c@{}}Mini-slot: to reducethe number of\\ symbols per TTI.\end{tabular}}                                                                                                                                                                                                                  & \multicolumn{1}{c|}{\multirow{2}{*}{Use}}                                                       & \multicolumn{2}{c|}{1}                                                                                & \multicolumn{2}{c|}{1}                                                                          & \multirow{2}{*}{Data}                                                                                     \\ \cline{3-5} \cline{7-10}
		& \multicolumn{1}{c|}{}                                                                                                                   & \multicolumn{3}{c|}{Based on FBC theory.}                                                                                                                                                                                                                                                                                          & \multicolumn{1}{c|}{}                                                                           & \multicolumn{2}{c|}{Preamble}                                                                         & \multicolumn{2}{c|}{Data}                                                                       &                                                                                                           \\ \hline
		\multirow{5}{*}{\begin{tabular}[c]{@{}c@{}}Queuing\\ delay\end{tabular}}                 & \multicolumn{1}{c|}{\multirow{2}{*}{\begin{tabular}[c]{@{}c@{}}Average\\ queuing\\ delay\end{tabular}}}                                 & \multicolumn{1}{c|}{\multirow{2}{*}{\begin{tabular}[c]{@{}c@{}}Queuing\\ theory\end{tabular}}}                      & \multicolumn{2}{c|}{Markov queuing model}                                                                                                                                                                    & \multicolumn{1}{c|}{\multirow{2}{*}{\begin{tabular}[c]{@{}c@{}}Number\\ of times\end{tabular}}} & \multicolumn{2}{c|}{\multirow{2}{*}{1}}                                                               & \multicolumn{2}{c|}{\multirow{2}{*}{0}}                                                         & \multirow{2}{*}{0}                                                                                        \\ \cline{4-5}
		& \multicolumn{1}{c|}{}                                                                                                                   & \multicolumn{1}{c|}{}                                                                                               & \multicolumn{2}{c|}{\begin{tabular}[c]{@{}c@{}}Two-dimensional\\ Markov chain\end{tabular}}                                                                                                                  & \multicolumn{1}{c|}{}                                                                           & \multicolumn{2}{c|}{}                                                                                 & \multicolumn{2}{c|}{}                                                                           &                                                                                                           \\ \cline{2-11} 
		& \multicolumn{1}{c|}{\multirow{3}{*}{\begin{tabular}[c]{@{}c@{}}Upper-bound\\ of queuing\\ delay\end{tabular}}}                          & \multicolumn{1}{c|}{\multirow{3}{*}{\begin{tabular}[c]{@{}c@{}}Large\\ deviation\\ principle\\ (LDP)\end{tabular}}} & \multicolumn{1}{c|}{\multirow{3}{*}{\begin{tabular}[c]{@{}c@{}}Based on\\ the concept\\ of effective\\ capacity.\end{tabular}}} & \begin{tabular}[c]{@{}c@{}}Resource\\ allocation\end{tabular}              & \multicolumn{1}{c|}{\multirow{3}{*}{Situation}}                                                 & \multicolumn{2}{c|}{\multirow{3}{*}{\begin{tabular}[c]{@{}c@{}}Queue\\ is not\\ empty.\end{tabular}}} & \multicolumn{2}{c|}{\multirow{3}{*}{\begin{tabular}[c]{@{}c@{}}Queue is\\ empty.\end{tabular}}} & \multirow{3}{*}{\begin{tabular}[c]{@{}c@{}}Immediate data\\ transmission\\ without waiting.\end{tabular}} \\ \cline{5-5}
		& \multicolumn{1}{c|}{}                                                                                                                   & \multicolumn{1}{c|}{}                                                                                               & \multicolumn{1}{c|}{}                                                                                                           & \begin{tabular}[c]{@{}c@{}}Scheduling\\ scheme\end{tabular}                & \multicolumn{1}{c|}{}                                                                           & \multicolumn{2}{c|}{}                                                                                 & \multicolumn{2}{c|}{}                                                                           &                                                                                                           \\ \cline{5-5}
		& \multicolumn{1}{c|}{}                                                                                                                   & \multicolumn{1}{c|}{}                                                                                               & \multicolumn{1}{c|}{}                                                                                                           & \begin{tabular}[c]{@{}c@{}}Heterogeneous\\ QoS\\ provisioning\end{tabular} & \multicolumn{1}{c|}{}                                                                           & \multicolumn{2}{c|}{}                                                                                 & \multicolumn{2}{c|}{}                                                                           &                                                                                                           \\ \hline
		\multirow{3}{*}{\begin{tabular}[c]{@{}c@{}}Processing\\ delay\end{tabular}}              & \multicolumn{4}{c|}{Mobile edge computing (MEC) for intelligent and fast responses}                                                                                                                                                                                                                                                                                                                                                                                          & \multicolumn{1}{c|}{\begin{tabular}[c]{@{}c@{}}Number\\ of times\end{tabular}}                  & \multicolumn{4}{c|}{3}                                                                                                                                                                                  & 1                                                                                                         \\ \cline{2-11} 
		& \multicolumn{4}{c|}{\begin{tabular}[c]{@{}c@{}}Distributed caching/control: to reduce redundant data\\ traffic and user content access delay.\end{tabular}}                                                                                                                                                                                                                                                                                                                  & \multicolumn{1}{c|}{\multirow{2}{*}{Location}}                                                  & \multicolumn{2}{c|}{2}                                                                                & \multicolumn{2}{c|}{1}                                                                          & \multirow{2}{*}{At BS}                                                                                    \\ \cline{2-5} \cline{7-10}
		& \multicolumn{4}{c|}{\begin{tabular}[c]{@{}c@{}}Parallel-processing strategy: to improve the efficiency\\ of cache servers.\end{tabular}}                                                                                                                                                                                                                                                                                                                                     & \multicolumn{1}{c|}{}                                                                           & \multicolumn{2}{c|}{At BS}                                                                            & \multicolumn{2}{c|}{At user}                                                                    &                                                                                                           \\ \hline
		\multirow{5}{*}{\begin{tabular}[c]{@{}c@{}}Propagation\\ delay\end{tabular}}             & \multicolumn{4}{c|}{Non-terrestrial networks for global coverage}                                                                                                                                                                                                                                                                                                                                                                                                            & \multicolumn{1}{c|}{\begin{tabular}[c]{@{}c@{}}Number\\ of times\end{tabular}}                  & \multicolumn{4}{c|}{4}                                                                                                                                                                                  & 2                                                                                                         \\ \cline{2-11} 
		& \multicolumn{4}{c|}{\multirow{2}{*}{Prediction and communication co-design for long-distance URLLC}}                                                                                                                                                                                                                                                                                                                                                                         & \multicolumn{1}{c|}{\multirow{4}{*}{Use}}                                                       & \multicolumn{2}{c|}{1}                                                                                & \multicolumn{2}{c|}{1}                                                                          & 1                                                                                                         \\ \cline{7-11} 
		& \multicolumn{4}{c|}{}                                                                                                                                                                                                                                                                                                                                                                                                                                                        & \multicolumn{1}{c|}{}                                                                           & \multicolumn{2}{c|}{Preamble}                                                                         & \multicolumn{2}{c|}{\begin{tabular}[c]{@{}c@{}}Access\\ response\end{tabular}}                  & Data                                                                                                      \\ \cline{2-5} \cline{7-11} 
		& \multicolumn{4}{c|}{\multirow{2}{*}{\begin{tabular}[c]{@{}c@{}}Non-geostationary (NGSO) satellite constellations for\\ satellite communications\end{tabular}}}                                                                                                                                                                                                                                                                                                               & \multicolumn{1}{c|}{}                                                                           & \multicolumn{2}{c|}{1}                                                                                & \multicolumn{2}{c|}{1}                                                                          & 1                                                                                                         \\ \cline{7-11} 
		& \multicolumn{4}{c|}{}                                                                                                                                                                                                                                                                                                                                                                                                                                                        & \multicolumn{1}{c|}{}                                                                           & \multicolumn{2}{c|}{Data}                                                                             & \multicolumn{2}{c|}{\begin{tabular}[c]{@{}c@{}}Contention\\ resolution\end{tabular}}            & \begin{tabular}[c]{@{}c@{}}Access\\ response\end{tabular}                                                 \\ \hline
		\multirow{8}{*}{\begin{tabular}[c]{@{}c@{}}Retransmission\\ times\end{tabular}}          & \multicolumn{4}{c|}{Access strategy: to reduce RA conflict.}                                                                                                                                                                                                                                                                                                                                                                                                                 & \multicolumn{1}{c|}{\multirow{5}{*}{\begin{tabular}[c]{@{}c@{}}Number\\ of times\end{tabular}}} & \multicolumn{4}{c|}{\multirow{5}{*}{Low}}                                                                                                                                                               & \multirow{5}{*}{High}                                                                                     \\ \cline{2-5}
		& \multicolumn{1}{c|}{\multirow{4}{*}{Reliability}}                                                                                       & \multicolumn{3}{c|}{Packet duplication}                                                                                                                                                                                                                                                                                            & \multicolumn{1}{c|}{}                                                                           & \multicolumn{4}{c|}{}                                                                                                                                                                                   &                                                                                                           \\ \cline{3-5}
		& \multicolumn{1}{c|}{}                                                                                                                   & \multicolumn{3}{c|}{\begin{tabular}[c]{@{}c@{}}Expanded channel state information (CSI)\\ feedback\end{tabular}}                                                                                                                                                                                                                   & \multicolumn{1}{c|}{}                                                                           & \multicolumn{4}{c|}{}                                                                                                                                                                                   &                                                                                                           \\ \cline{3-5}
		& \multicolumn{1}{c|}{}                                                                                                                   & \multicolumn{3}{c|}{Slot aggregation}                                                                                                                                                                                                                                                                                              & \multicolumn{1}{c|}{}                                                                           & \multicolumn{4}{c|}{}                                                                                                                                                                                   &                                                                                                           \\ \cline{3-5}
		& \multicolumn{1}{c|}{}                                                                                                                   & \multicolumn{3}{c|}{\begin{tabular}[c]{@{}c@{}}Multiple transmission and reception\\ points (TRPs) support\end{tabular}}                                                                                                                                                                                                           & \multicolumn{1}{c|}{}                                                                           & \multicolumn{4}{c|}{}                                                                                                                                                                                   &                                                                                                           \\ \cline{2-11} 
		& \multicolumn{1}{c|}{\multirow{3}{*}{\begin{tabular}[c]{@{}c@{}}Hybrid automatic\\ repeat request\\ (HARQ)\\ enhancements\end{tabular}}} & \multicolumn{3}{c|}{\begin{tabular}[c]{@{}c@{}}Flexible HARQ round triptimes (RTT)\\ configuration\end{tabular}}                                                                                                                                                                                                                   & \multicolumn{1}{c|}{\multirow{3}{*}{Resolution}}                                                & \multicolumn{4}{c|}{\multirow{3}{*}{Contention-free}}                                                                                                                                                   & \multirow{3}{*}{Contention-based}                                                                         \\ \cline{3-5}
		& \multicolumn{1}{c|}{}                                                                                                                   & \multicolumn{3}{c|}{Low-order modulation and coding scheme}                                                                                                                                                                                                                                                                        & \multicolumn{1}{c|}{}                                                                           & \multicolumn{4}{c|}{}                                                                                                                                                                                   &                                                                                                           \\ \cline{3-5}
		& \multicolumn{1}{c|}{}                                                                                                                   & \multicolumn{3}{c|}{\begin{tabular}[c]{@{}c@{}}Adaptive redundancy matching with\\ enriched feedback\end{tabular}}                                                                                                                                                                                                                 & \multicolumn{1}{c|}{}                                                                           & \multicolumn{4}{c|}{}                                                                                                                                                                                   &                                                                                                           \\ \hline
	\end{tabular}
\end{table*}

\subsection{Access Protocol-Specified Delay}
In real scenarios, since the retransmission times, specific delay components, and delay values are related to different access protocols, it is necessary to analyze the access protocols to obtain the specific delay of real-time services. In Table~\ref{Table:Two_Table}, we provide the specific access delay composition and comparison of GB and GF RA protocols.

\subsubsection{The Delay in GB RA}
Table~\ref{Table:Two_Table} shows the access delay components in GB RA. The scheduling request (preamble) and contention resolution procedure in GB RA lead to an additional access delay of at least several milliseconds within the Long-Term Evolution (LTE) network configurations, which is unsuitable for low-latency communications. 

\subsubsection{The Delay in GF RA}
Due to the characteristic of intermittent and bursty traffic regarding delay-sensitive tasks, GF RA is proposed as an efficient transmission technology to support low-latency connectivity. As shown in Table~\ref{Table:Two_Table}, in GF RA, unlike GB RA, active users are not required to wait for a grant from the BS, which significantly reduces delay by eliminating the need for the transmission grant process. This means that as soon as a user becomes active, it immediately sends data along with a preamble and awaits acknowledgment (ACK) from the BS, thereby bypassing the grant step and consequently reducing delay~\cite{GF}.

\subsection{Existing Problems}

\subsubsection{Separate Delay Components Optimization} 
For the conventional over-the-air delay optimization problem, each delay component is separately optimized as an independent part as shown in Table~\ref{Table:Two_Table}. Optimizing a single component of the end-to-end delay separately lacks systematic and comprehensive consideration.

\subsubsection{Fixed Blocklength Structure} 
The TTI design of LTE and 5G NR are both fixed, which limits the flexibility of system design for transmission delay reduction. When heterogeneous packets with different delay requirements and packet sizes arrive at the same time, fixed frame structure is difficult to adapt to variable delay and traffic situations.

\subsubsection{Lack of Specific Access Protocol Combined Analysis} 
Different access protocols relate to different kinds of composition regarding transmission delay, queuing delay, processing delay, and propagation delay, as well as different retransmission times. However, existing research lacks delay analysis combined with specific access protocols, which cannot demonstrate the overall end-to-end delay performance of the system.

\section{Delay Tradeoff Relationship\\in The FBL Regime}
To solve the above problems, we delve into the delay tradeoff relationship in the FBL regime. Based on the newly shown tradeoff, new system delay performance can be derived in the FBL regime and the new frame structure combined with specific access protocols can be accordingly redesigned.

\subsection{The Structure And Impact of FBL}
The blocklength refers to the number of symbols transmitted within a block, and it is determined by multiplying the time span (TTI) and the subchannel bandwidth resource. Short packets in URLLC correspond to finite blocklengths, and the corresponding achievable rate in the FBL regime has been derived. It is a function of blocklength and decreases as the blocklength decreases~\cite{ChannelRate}. Especially in areas where the blocklength is short, the achievable rate is severely impacted. Different from the traditional long-frame structure, in the case of short packets with short-frame structure, the influence of FBL on system performance should be fully considered.

\subsection{Tradeoff Relationship Among Transmission Delay, Queuing Delay, and Retransmission Times}

\begin{figure}[htbp]
	\centering
	\includegraphics[scale =0.72]{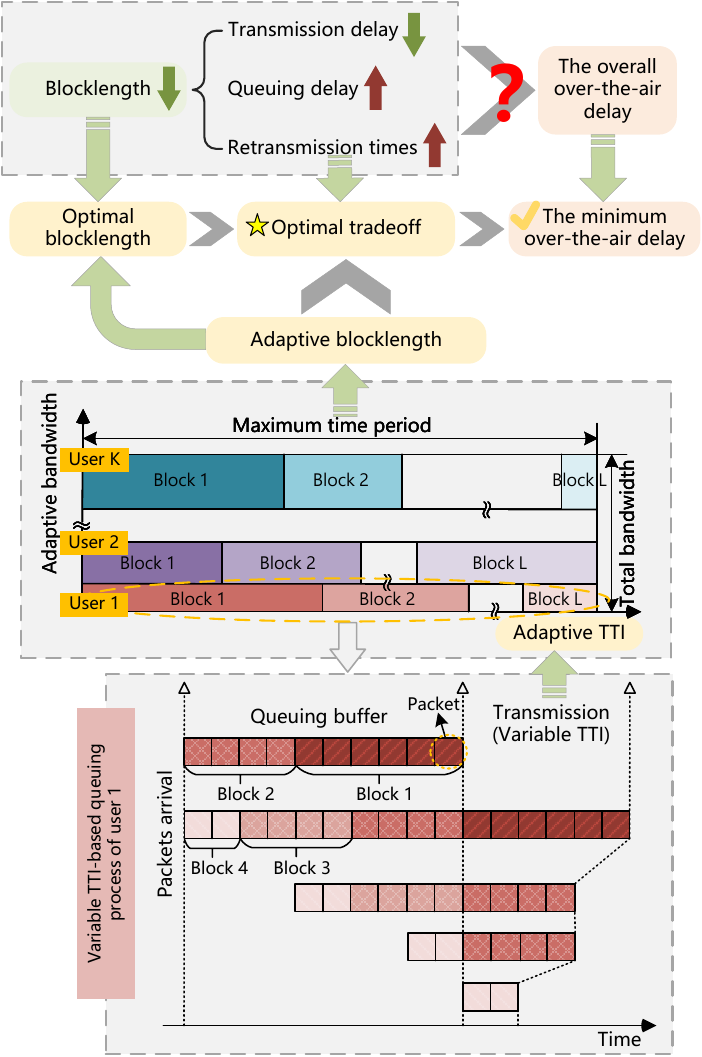}
	\caption{Tradeoff relationship, adaptive blocklength framework, and variable TTI-based dynamic queuing process.}
	\label{Tradeoff-Adaptive}
\end{figure}

\subsubsection{IBL Regime}
In the traditional IBL regime, the achievable rate remains constant regardless of the blocklength. As a result, the service rate of the queuing process in the link layer does not change, leading to an unchanged queuing delay. Furthermore, since the transmission error probability approaches zero in the IBL regime, the variation in blocklength does not affect retransmission times. Only the transmission delay changes with the blocklength, decreasing as the blocklength decreases. Consequently, the over-the-air delay also monotonously decreases with the reduction in blocklength.

\subsubsection{FBL Regime}
Fig.~\ref{Tradeoff-Adaptive} shows the change of transmission delay, queuing delay, and retransmission times with the finite blocklength. In the FBL regime, a reduction in blocklength leads to a decrease in transmission delay. However, this reduction in blocklength also leads to a decrease in the achievable rate, which in turn causes an increase in queuing delay. Besides, unlike the assumption in the IBL regime, the FBL leads to additional error probability, which no longer approaches zero. As the blocklength decreases, the error probability increases, leading to more frequent retransmissions. 

The transmission delay, queuing delay, and retransmission times do not change in the same direction with the blocklength, which reveals a tradeoff relationship among them. The minimum over-the-air delay can be achieved when the optimal tradeoff is reached, and the corresponding blocklength is the optimal blocklength. In order to achieve the optimal tradeoff for minimizing over-the-air delay, the following aspects should be taken into account.

\subsection{Factors for Achieving Optimal Tradeoff}
In order to obtain the optimal tradeoff relationship to reduce the over-the-air delay, the following three aspects should be considered.

\subsubsection{Adaptive Blocklength Framework} 
Even the scalable slot based scheduling can simultaneously support a variety of applications, the TTI is still fixed, which prevents further reduction of the over-the-air delay. Based on the above analysis, the change of blocklength per frame impacts the over-the-air delay. Under different packet arrival rates, packet sizes, and traffic loads, the optimal tradeoff among transmission delay, queuing delay, and retransmission times cannot be achieved with a fixed blocklength-based frame structure. This situation results in an increase in the overall over-the-air delay. Therefore, it is necessary to study a new frame framework with adaptive blocklength to find out the optimal blocklength.

\subsubsection{Joint Optimization of Different Delay Components}
Based on the tradeoff analysis, the optimization of a single delay component may lead to the increase of other delay components. Thus, each delay part should be combined with other delay components to form a comprehensive delay analysis and joint optimization method. To achieve the lowest over-the-air delay, it is necessary to jointly optimize different delay components to find the optimal global solution instead of treating them as independent parts.

\subsection{Applicability for Short And Long Packets Coexistence} 
In different types of low-latency tasks, real-time interactions are operated using text, speech, images, and augmented/virtual reality, which relate to different packet types, packet sizes and delay requirements. Based on real-time packet arrival rates and task loads, the proposed adaptive blocklength framework can flexibly adjust the blocklength for both short and long packets to accommodate the diverse delay requirements of various tasks. The specific implementation strategy is in the following.
\subsubsection{Traffic Monitoring and Analysis}
Deploy traffic monitoring systems to analyze the characteristics of real-time traffic, including the size of packets, transmission frequency, and service type. Based on the results of traffic analysis, dynamically adjust the blocklengths of packets. 

\subsubsection{Performance Testing and Feedback Mechanism}
Establish a real-time feedback mechanism to adjust the strategy according to changes in network conditions, ensuring that the network operates optimally and enabling assessment and further optimization of the improvements based on test results.

By adaptively adjusting the blocklengths of packets, better QoS can be provided according to the needs of different applications, especially for applications sensitive to delays.

\section{A Case Study: Adaptive Blocklength Scheme for URLLC}
In this section, we propose a dynamic queuing process based on variable TTI and the adaptive blocklength framework. Then, we carry out a case study to solve the problem of minimizing the over-the-air delay.

\subsection{Variable TTI-based Dynamic Queuing Process}
In contrast to the fixed TTI structure in 5G NR, we set a variable TTI for each packet, as illustrated in Fig.~\ref{Tradeoff-Adaptive}. Depending on the packet arrival rates and traffic load conditions, different TTIs are set when the packets are in the queuing buffer, and the packet transmission stage also features different TTIs. The TTIs are considered to be changed, where the corresponding blocklengths are also changed. 

\subsection{Adaptive Blocklength Framework in The FBL Regime for URLLC}
To achieve the optimal tradeoff relationship among transmission delay, queuing delay, and retransmission times, we consider optimizing the blocklength of each frame. Given that the blocklength is the product of TTI and bandwidth, optimizing blocklength is equivalent to optimizing time-bandwidth resource allocation. In this case, it is necessary to jointly adjust these two factors to create the corresponding adaptive blocklength. That is, in addition to the above variable TTI, we also allocate bandwidth, that is, we assign subchannels to different users. 

Fig.~\ref{Tradeoff-Adaptive} illustrates the framework for adaptive blocklength, where each packet has different TTI and bandwidth configuration. In the frequency domain, the total bandwidth is allocated to all users, thus forming adaptive bandwidth. Similarly, in the time domain, the maximum time period is divided into several parts, allocated to packets of each user, thereby creating adaptive TTIs. Through the adaptive TTI and bandwidth, the blocklength of each frame can be adaptively changed, where the adaptive blocklength framework is formed. In addition to the above framework, which contributes to the improvement of delay performance, we set a fixed threshold for transmission error to simultaneously ensure high reliability in URLLC.

\subsection{Singer-User Or NOMA Scenarios}
\subsubsection{Adaptive TTI-Based Over-the-Air Delay Minimization Problem}
To minimize the over-the-air delay, based on the revealed tradeoff relationship, we consider optimizing the blocklength of each frame to balance these three delay components. In single-user or NOMA scenarios, a fixed bandwidth is adopted without loss of generality, where the bandwidth resources have been normalized. This means that adaptive blocklength is equal to adaptive TTI.

We set the average over-the-air delay as the objective function of the problem. Then, we set up three constraints. First, the total number of bits contained in serviced packets needs to be transmitted within the current period to ensure reliability. Second, the sum of blocklength for all frames cannot exceed the period duration to avoid packet loss. Third, the blocklength of each frame should be set larger than or equal to zero. The variable we optimize is the blocklength of each frame. Using the adaptive blocklength framework, we can dynamically adjust the blocklength to reduce the over-the-air delay.

\subsection{Multi-User Scenario with OMA Scenarios}
\subsubsection{Joint Adaptive Bandwidth and TTI-Based Average Over-the-Air Delay Minimization Problem}
Different from the single-user or NOMA scenarios, multiple users share multiple subchannels. According to the relationship between blocklength and time-bandwidth resource, optimizing blocklength is equivalent to optimizing time-bandwidth resource allocation. Therefore, in the multi-user with OMA scenario, an additional constraint needs to be added that the bandwidth resource of each subchannel can only be allocated to one user.

Besides, the corresponding problem in the multi-user with OMA scenario is more complicated due to the discreteness of bandwidth allocation brought by subchannels. Thus, this average over-the-air delay minimization problem includes continuous non-convex and discrete constraints, which cannot be solved using the traditional mathematical methods. A series of deep reinforcement learning (DRL)-based algorithms can be used to tackle this non-convex problem. Since the off-line training process is done before being implemented into the system, it can be applied to low-latency communication systems.

\subsubsection{DDPG} Based on the algorithm idea of deterministic policy gradient (DPG) and deep learning, deep deterministic policy gradient (DDPG), with dual network and experience replay, can solve the problem that the DRL network is difficult to converge~\cite{DDPG}. DDPG can be used to solve the adaptive bandwidth and TTI-based delay minimization problem in the multi-user scenario.

\subsubsection{Multi-DQN} Two subproblems need to be solved here, one is the bandwidth allocation problem, and the other is the TTI design problem. Even a small number of users, subchannels and frames can lead to a large action space. The curse of dimensionality in the action space is a serious problem in the deep Q-network (DQN) since the exponentially increasing action space makes DQN quite complex and sample inefficient, where the corresponding convergence speed will be slowed down. Thus, Multi-DQN is used to decompose the large action space. Using two DQNs to allocate bandwidth and design TTI respectively, Multi-DQN can avoid large action space and low convergence speed.

\subsubsection{Performance Comparison of DDPG And Multi-DQN}
\begin{figure}[htbp]
	\centering
	\includegraphics[scale = 0.49]{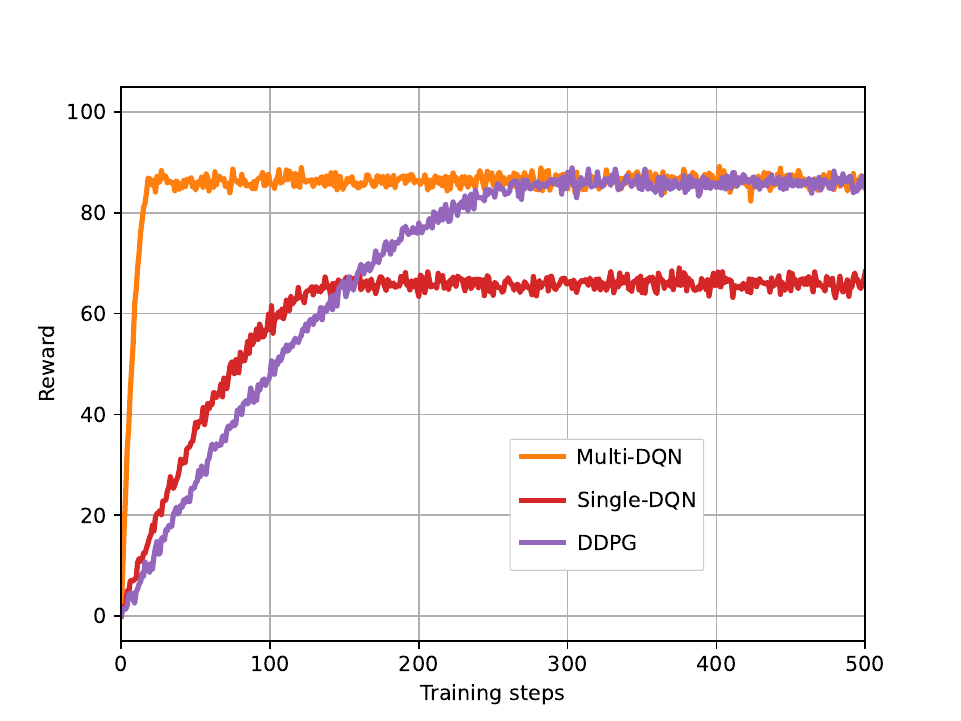}
	\caption{The convergence comparison of DDPG, Multi-DQN, and Single-DQN.}
	\label{fig:DDPG_MultiDQN_SingleDQN}
\end{figure}

Figure~\ref{fig:DDPG_MultiDQN_SingleDQN} shows the convergence comparison of DDPG, Multi-DQN, and Single-DQN for multi-user adaptive blocklength scheme. As illustrated in Fig.~\ref{fig:DDPG_MultiDQN_SingleDQN}, the reward of Single-DQN does not reach those of Multi-DQN and DDPG. Multi-DQN is capable of matching the reward of DDPG, indicating that Multi-DQN can achieve the performance limit set by DDPG. Additionally, Multi-DQN converges faster than DDPG, with the convergence time for Multi-DQN being approximately $1/10$ of that for DDPG. This means Multi-DQN can degrade the dimension of action space and fast converge to optimal solution compared with Single-DQN and DDPG, implying reduced energy consumption for calculation is required due to fewer iterations in Multi-DQN. For future URLLC and IoT scenarios, the low power consumption characteristic of Multi-DQN can reduce the energy consumption caused by a large number of sensors and connectivities. Therefore, we choose Multi-DQN to achieve the adaptive blocklength scheme.

\subsection{Delay Performance Evaluation}
We consider the URLLC scenario, where packets arrive with a size of $32$ bytes. The maximum transmit power per user is set to $100$~mW and the total bandwidth is set to $1$~MHz. The channel gain for each user is set in the range of $[-20\rm dB, 20\rm dB]$. In order to satisfy the high-reliability requirements in URLLC, the transmission error probability is set to $10^{-7}$. In addition, the duration of one block in 5G NR is set as $0.5$~ms, $0.25$~ms, $0.125$~ms, and $0.0625$~ms, respectively. The duration of one block in LTE is set as $1$~ms.
\begin{figure}[htbp]
	\centering
	\subfigure[Delay performance comparison versus total time.]{
		\includegraphics[width=6.9cm]{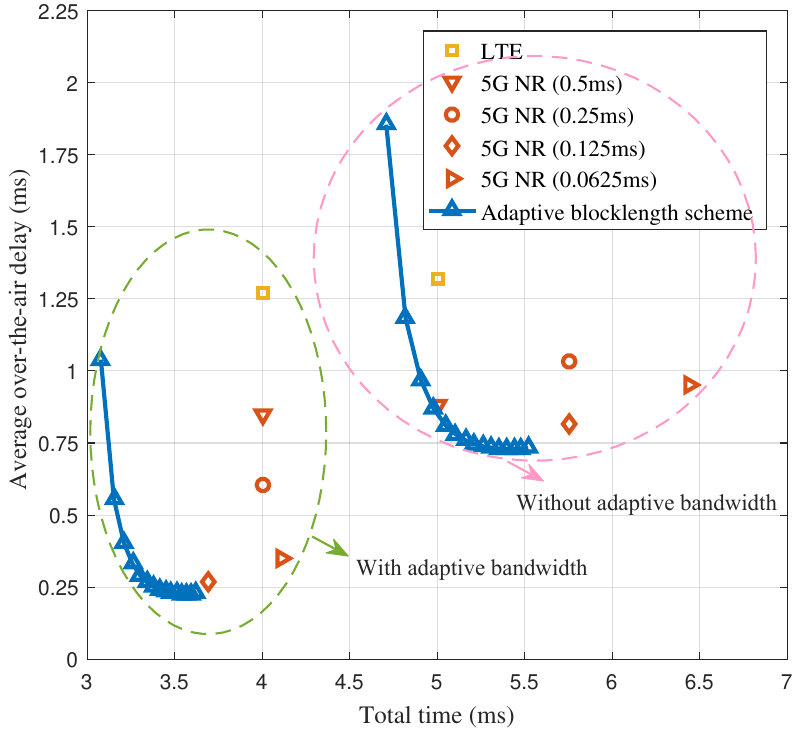}
	}
	\subfigure[Delay performance comparison versus the number of users.]{
		\includegraphics[width=6.9cm]{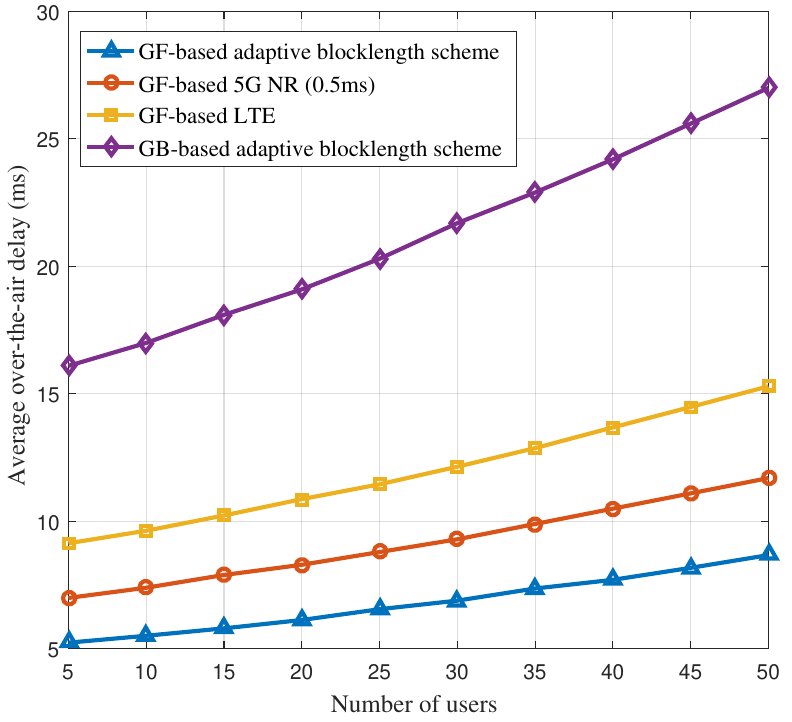}
		
	}
	\centering \caption{Delay performance comparison.}\label{fig:simulation}
\end{figure}

Figure~\ref{fig:simulation}(a) depicts the delay performance of the proposed adaptive blocklength framework and the fixed blocklength structures in LTE and 5G NR. Two performance indicators are evaluated, namely the average over-the-air delay and the total time, where the total time is the time consumption to transmit all packets. The average over-the-air delay and the total time of the proposed adaptive blocklength framework are both less than those in LTE and 5G NR. Also, Fig.~\ref{fig:simulation}(a) demonstrates that if only the blocklength adaptively changes with fixed bandwidth, the corresponding average over-the-air delay is longer than that of LTE and 5G NR with adaptive bandwidth allocation, indicating the importance of adaptive bandwidth allocation in our proposed framework.

Figure~\ref{fig:simulation}(b) shows the delay performance comparison in GF RA and GB RA under different numbers of users. The over-the-air delay of GB-based scheme is significantly higher than that of GF-based scheme, which means GF RA is more suitable for URLLC. With the same number of users, GF-based adaptive blocklength schemes can attain the lowest average over-the-air delay compared to LTE and 5G NR. And when the number of users changes, the blocklength can be dynamically adjusted to accommodate real-time tasks.

\section{Conclusions}
In this article, a new delay reduction framework, namely the adaptive blocklength framework, was proposed for URLLC to support delay-sensitive services. By comprehensively analyzing related components of the end-to-end delay, the FBL transmission theory, and delay combined with the access protocol, we obtained the tradeoff relationship among transmission delay, queuing delay, and retransmission times. On this basis, we developed the adaptive blocklength framework and corresponding case study, which can help to achieve the optimal balance of the tradeoff. Compared to the fixed blocklength structures in LTE and 5G NR, our proposed framework can reduce the over-the-air delay and increase flexibility in system design. We hope this article guides the researchers and practitioners interested in applying the adaptive blocklength framework to related issues for URLLC.

\bibliographystyle{IEEEtran}
\bibliography{References}

\end{document}